\documentclass[sn-nature]{sn-jnl}% Style for submissions to Nature Portfolio journals

\usepackage[T1]{fontenc}
\usepackage[latin9]{inputenc}
\usepackage{graphicx}%
\usepackage{float}
\usepackage{multirow}%
\usepackage{amsmath,amsthm,mathrsfs,amsfonts,amssymb,bbm}   % for math
\usepackage[title]{appendix}%
\usepackage{xcolor}%
\usepackage{textcomp}%
\usepackage{manyfoot}%
\usepackage{booktabs}%
\usepackage{bm}
\usepackage{framed}
\usepackage{braket}
\usepackage{soul}

\usepackage{longtable}
\usepackage{esint}
\usepackage{babel}
\usepackage{afterpage,lscape}

\usepackage{hyperref}
\usepackage{cleveref}
\usepackage{anyfontsize}
\renewcommand{\tablename}{Tab.}

\begin{document}

\title[Article Title]{Half-Quantized Hall Metal and Marginal Metal
in Disordered Magnetic Topological Insulators}

\author[1]{\fnm{Bi} \sur{Shi-Hao}}
\author*[2]{\fnm{Fu} \sur{Bo}}\email{fubo@gbu.edu.cn}
\author*[1]{\fnm{Shen} \sur{Shun-Qing}}\email{sshen@hku.hk}

\affil[1]{\orgdiv{Department of Physics}, \orgname{The University of Hong Kong}, \orgaddress{\street{Pokfulam Road}, \city{Hong Kong},   \country{China}}}
\affil[2]{\orgdiv{School of Sciences}, \orgname{Great Bay University}, \orgaddress{\city{Dongguan}, \postcode{523000}, \state{Guangdong Province}, \country{China}}}

\abstract{A semimagnetic topological insulator---a heterostructure combining
a topological insulator with a ferromagnet---exhibits a half-quantized
Hall effect, characterized by a quantized Hall conductance of $\frac{1}{2}\frac{e^{2}}{h}$
(where $e$ is the elementary charge and $h$ is the Planck constant), 
which reinforces the established understanding of topological phenomena in condensed matter physics.
However, its stability in realistic, disordered systems remains poorly understood.
Here, we demonstrate the robustness of the half-quantized Hall effect in weakly disordered systems, 
stemming from a single gapless Dirac cone of fermions and coexisting with weak antilocalization due to the
$\pi$ Berry phase that suppresses backscattering. 
Furthermore, we uncover a marginal metallic phase emerging between weak antilocalization and 
Anderson insulation---a transition that defies conventional metal-insulator transitions by 
lacking an isolated critical point---where both conductance and normalized localization length 
exhibit scale invariance, independent of system size.
The half-quantized Hall metal and the marginal metallic phase challenge existing localization theories 
and provide insights into disorder-driven topological phase transitions in magnetic topological insulators, 
opening avenues for exploring quantum materials and next-generation electronic devices.}

\maketitle

\section*{Introduction}\label{sec:introduction}

The quantum anomalous Hall effect (QAHE) in ferromagnetic insulators
is a cornerstone of modern condensed matter physics, characterized
by quantized Hall conductivity in the absence of an external magnetic
field, arising from the topological properties of electronic states
\cite{Nagaosa-10rmp,xiao2010berry,ChangCZ-23rmp}. In two-dimensional
metallic ferromagnets, the Hall conductivity is typically non-quantized
and can be expressed as an integral of Berry connection over the Fermi
surface \cite{haldane2004berry}. Recent theoretical and experimental
advances have revealed that systems featuring a single gapless Dirac
cone of electrons in the first Brillouin zone can exhibit a half-quantized
Hall conductivity \cite{Fu2022:npjQM,Zou2022:PRB,Zou2023:PRB}. This
phenomenon bears a striking resemblance to the parity anomaly of massless
Dirac fermions in quantum field theory \cite{redlich1984parity,Semenoff1984:PRL,qi2011topological,Shen-24coshare}.
Unlike the QAHE observed in insulating phases, which is characterized
by integer Chern numbers and the emergence of chiral edge states \cite{thouless1982quantized,Haldane1988:PRL,yu2010quantized,Chu2011:PRB,chang2013experimental,checkelsky2014trajectory,XXD2023:Nat1,XXD2023:Nat2,lu2024fractional},
the half-quantized Hall effect (HQHE) occurs in systems with a finite
Fermi surface and non-zero longitudinal conductivity. The effect has
been experimentally observed in a semi-magnetic structure of Cr-doped
topological insulator (TI) $\mathrm{(Bi,Sb)_{2}\mathrm{Te}_{3}}$
\cite{Mogi2022:NatPhys}, sparking significant interest in its realization,
robustness, and dissipative properties \cite{gong2023half,yang2023realization,ning2023robustness,wang2024signature,wan2024quarter,zhou2024dissipative}.

Despite these advances, the HQHE presents a fundamental challenge
to the theory of localization. In two-dimensional systems with broken
time-reversal symmetry, disorder-induced localization of electronic
states is commonly anticipated, leading to integer-quantized Hall
conductivity \cite{Pruisken}. This raises a critical question: Can
the HQHE in semimagnetic topological insulators (TIs) survive or remain
robust in the presence of disorder? The localization behavior of disordered
systems depends on their dimensionality and the symmetry class of
the Hamiltonian \cite{Mirlin2008:RMP}. Recent insights have highlighted
the profound influence of topological wave functions on localization
properties \cite{Konig2012:PRB,Mirlin2023:PRB,Ostrovsky2007:PRL,Bardarson:PRL,Nomura2007:PRL,Shindou2023:PRL,Shindou2024:PRL}.
Specifically, relativistic Dirac electrons, such as those in a single-flavor
gapless Dirac cone, are theoretically resistant to localization even
under strong disorder \cite{Ostrovsky2007:PRL,Nomura2007:PRL,Bardarson:PRL},
provided there is no coupling to other bands. This suggests that a
system exhibiting the HQHE should retain its metallic character under
moderate disorder. In condensed matter systems, a single-flavor Dirac
cone typically appears on one surface of a three-dimensional $\mathbb{Z}_{2}$
topological insulator, remaining robust until strong disorder drives
the system into a topologically trivial insulating state with zero
Hall conductivity \cite{Kobayashi2013:PRL}. Given that a semimagnetic
TI hosts only a single gapless Dirac cone, the transition between
the HQHE and the Anderson insulating state warrants thorough exploration.

In this work, we employ well-established numerical methods (see \hyperref[sec:Methods]{\textit{Methods}})
to investigate disorder-driven topological phase transitions in 
semimagnetic topological insulator thin films modeled on a real-space lattice.
We reveal the robustness of the disordered half-quantized Hall metal and the emergence 
of a marginal metallic phase, shedding light on topological materials, disorder physics, 
and quantum transport research. 
Moreover, since the half-quantized quantum Hall effect has been experimentally observed \cite{Mogi2022:NatPhys}, 
the phenomena we describe offer insights for future experimental design and observation 
of related effects.

\section*{Results}\label{sec:results}

\paragraph{The Phase Diagram}\label{subsec:Evolution}

\begin{figure}[H]
\centering
\includegraphics[width=12cm]{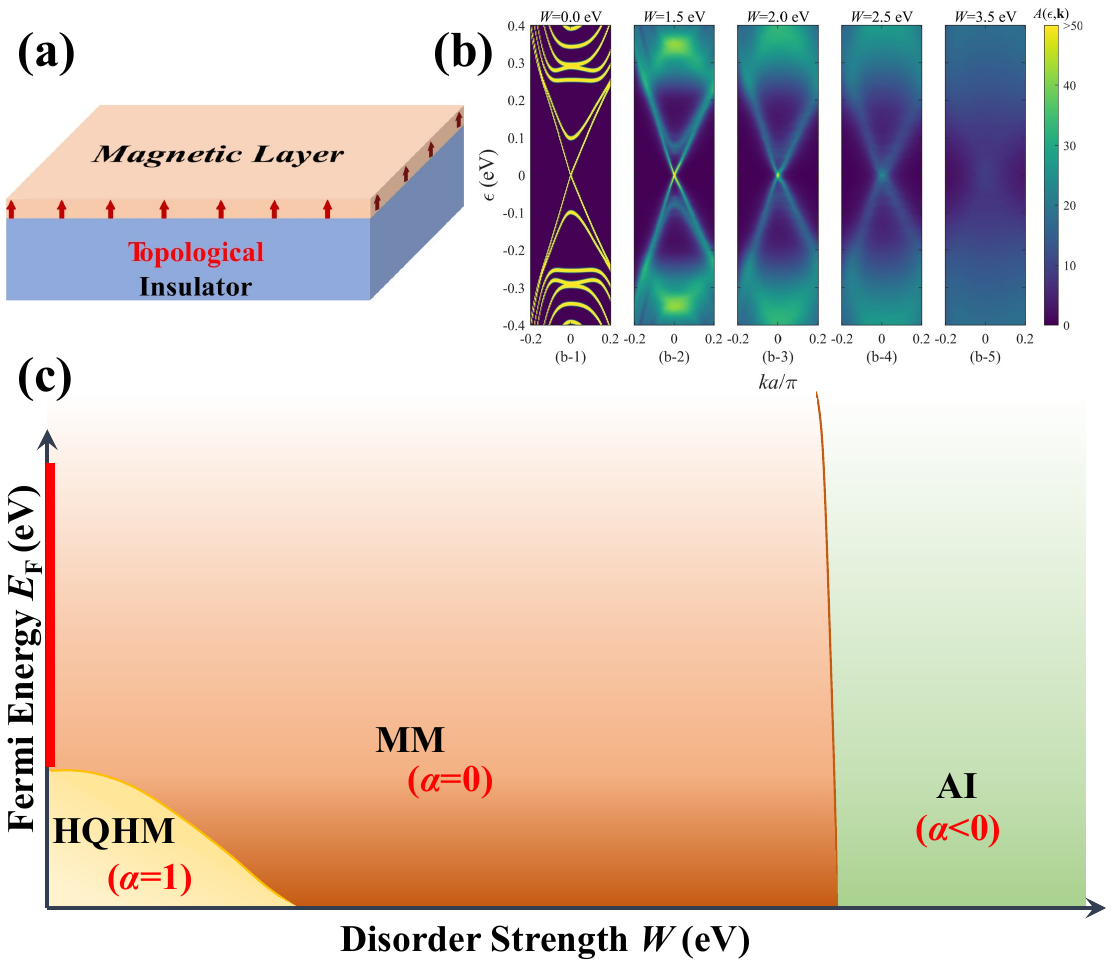}
\caption{\textbf{The phase diagram in disordered semimagnetic topological insulator
films.} (a). Schematic of a semimagnetic topological insulator film.
 The acronyms HQHM (Half-Quantized Hall Metal), MM (Marginal Metal), and AI (Anderson insulator) label distinct phases in the diagram. The red solid line indicates that the clean system
is a metal rather than MM. (b). Evolution of the spectral function
$A(\epsilon,\mathbf{k})$ with the disorder strength $W$.
(c). The phase diagram of the semimagnetic topological insulator in
the $W-E_{\rm F}$ plane: The metallic phases of the scaling function
$\alpha=1$ with the half-quantized anomalous Hall conductivity; the
marginal metallic phase of $\alpha=0$ with non-quantized Hall conductivity,
and the Anderson insulating phase. Here we take the system size $L$ large enough.}
\label{fig:phase1}
\end{figure}

The main results of this study are summarized in Fig. \ref{fig:phase1},
which depicts the phase diagram of a disordered semimagnetic topological
insulator (TI) in the $E_{\rm F}-W$ plane, where $E_{\rm F}$ is the chemical
potential and $W$ is the disorder strength. The phase diagram comprises
three distinct regions: (1) the half-quantized Hall metal
(HQHM) at weak disorder, (2) the marginal metal (MM) at
moderate disorder, and (3) the Anderson insulator (AI) at
strong disorder. In the absence of disorder, a semimagnetic TI hosts
a single gapless Dirac cone of fermions on one surface and a massive
Dirac cone of fermions on the other surface. The gapless Dirac cone
arises from the surface electrons of the underlying strong topological
insulator, localized on the bottom layer, which is spatially separated
from the top magnetic-doped layers. The presence of this single gapless
Dirac cone results in a half-quantized Hall conductance of $\frac{1}{2}\frac{e^{2}}{h}$
\cite{Zou2023:PRB,fu2025z}. In the weak disorder regime, the half-quantized
Hall conductance remains robust against disorder, defining the HQHM
phase. In this region, the suppression of backscattering for gapless
Dirac fermions leads to weak antilocalization (WAL) behavior in the
longitudinal conductivity $\sigma$. Specifically, the quantum correction
to the conductivity follows $\delta\sigma\simeq\frac{e^{2}}{\pi h}\ln(L/l_{e})$,
where $L$ is the sample size and $l_{e}$ is the elastic mean free
path. This logarithmic increase in conductivity with sample size indicates
a metallic phase. By introducing the scaling function $\alpha=\pi\frac{h}{e^{2}}\frac{\partial\sigma}{\partial\ln L}$,
the HQHM phase is characterized by $\alpha=1$ when
the Fermi energy only intersects with the gapless Dirac cone. As
the disorder strength increases, the system transitions into the MM 
phase \cite{xie1998kosterlitz,zhang2009localization,kettemann2009critical,zhang2024interlayer,wang2024anderson,wang2015band,chen2019metal,ChenCZ2019:PRL}
in which interband scattering among Dirac bands plays
a significant role. In this regime, the conductivity becomes scale-invariant,
exhibiting no dependence on system size ($\alpha=0$). Simultaneously,
the Hall conductivity decays gradually to vanish. Crucially, the MM
phase persists over a finite range of disorder strengths, in contrast
to conventional metal-insulator transitions, which occur at a single
critical point. This extended phase highlights the unique nature of
the disorder-driven transition in semimagnetic TIs. At even stronger
disorder, the system enters the AI phase, where all electronic states become fully localized.

This phase diagram reveals a disorder-driven transition pathway
in semimagnetic TIs, distinct from conventional metal-insulator transitions.
The robustness of the HQHM phase, the emergence of the scale-invariant
MM phase, and the eventual transition to the AI phase underscore the
intricate interplay between topology, disorder, and electronic localization
in these systems.

\paragraph*{Half-quantized Hall Effect in Weak Disorder}\label{subsec:HQHE}

\begin{figure}[H]
\centering
\includegraphics[width=12cm]{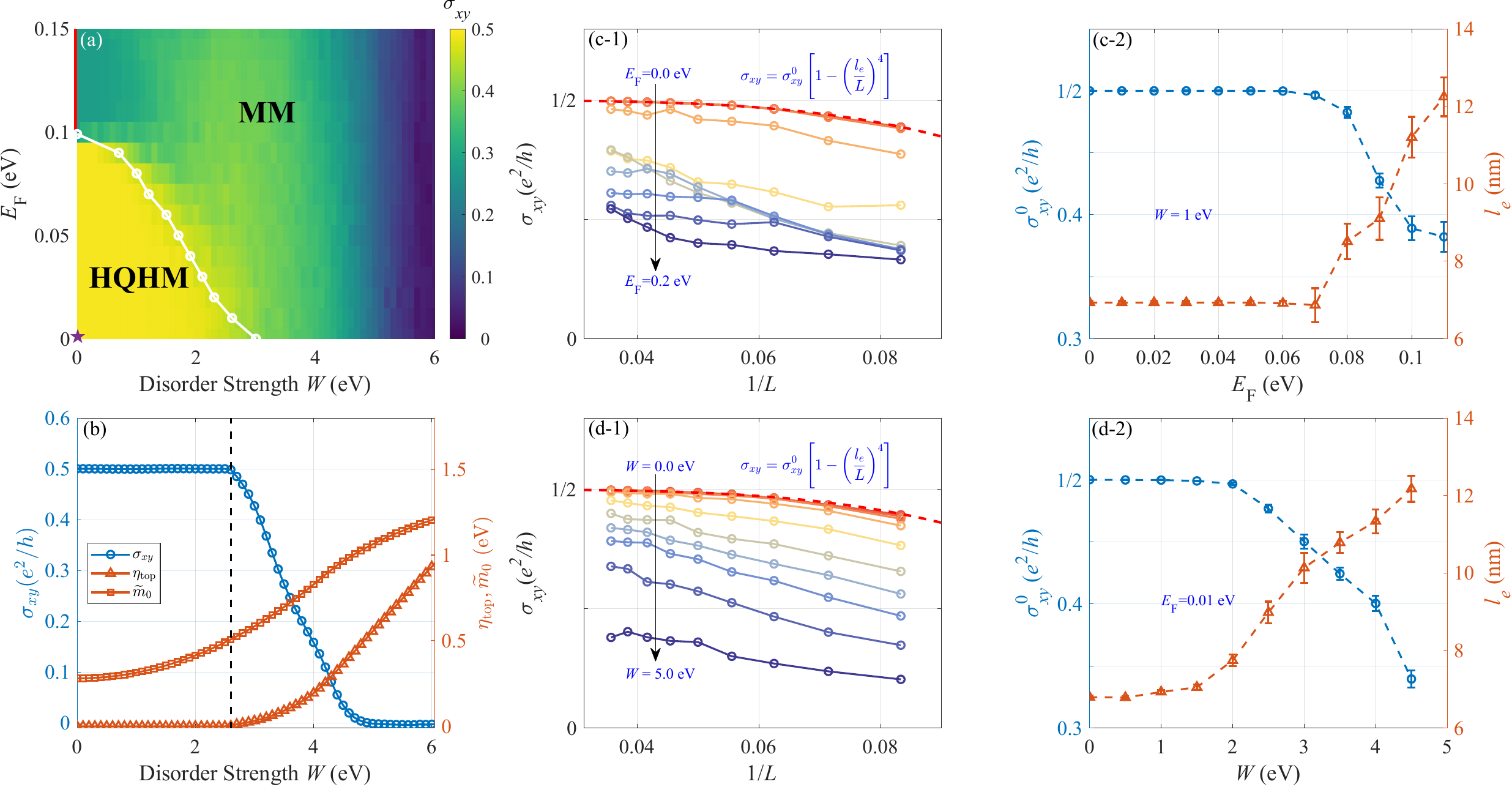}
\caption{\textbf{Robustness of the Half-quantized Hall effect in Weak Disorder.
} (a). The phase diagram of the Hall conductivity in the $W-E_{\rm F}$
plane. We set the lattice size $L_{x}=L_{y}=L=20$ in the simulation.
The red solid line indicates that the clean system
is metallic. The bright yellow areas highlight the half-quantized Hall metal (HQHM) phase, whereas
the chartreuse region indicates marginal metal (MM) with non-quantized Hall conductivity. The
white solid line marks the phase boundary as determined by the effective
medium theory. (b) The calculated Hall conductivity and disorder renormalized
Dirac mass $\widetilde{m}_{0}$ and energy broadening $\eta_{\mathrm{top}}$
at $E_{\rm F}=0.01$ eV as a function of $W$ in the effective medium
theory. The black dashed line denotes the critical threshold $W_{\mathrm{c}}=2.6$
eV. The finite-size scaling analysis of the quantization error, $\sigma_{xy}^{0}-\sigma_{xy}$,
is presented as the function of $1/L$ in (c-1) along the vertical
line of $W=1.0$ eV for varying $E_{\rm F}$, and (d-1) along the horizontal
line of $E_{\rm F}=0.01$ eV for varying $W$. The corresponding $\sigma_{xy}^{0}$ and $l_{e}$
are displayed in (c-2) and (d-2), respectively. The error bars reflect the uncertainties 
arising from the numerical fitting. We have used the set
of parameters $L_{z}=10$, $L_{z}^{\mathrm{Mag}}=3$, $\lambda_{\parallel}=0.41$
eV, $\lambda_{z}=0.44$ eV, $t_{\parallel}=0.566$ eV, $t_{z}=0.40$
eV, $V_{0}=0.1$ eV, and lattice constants $a=b=1$ nm and $c=0.5$
nm unless otherwise specified. The raw data points are averaged over
50 random samples.}
\label{fig:phase2}
\end{figure}

To demonstrate the robustness of the HQHE, we first calculate the
Hall conductivity $\sigma_{xy}$ for a tight-binding model of the
semimagnetic TI on an $L_{x}\times L_{y}\times L_{z}$ lattice using
the real-space Kubo formula \cite{Prodan2011:JPA} (see \hyperref[subsec:RSCN]{\textit{Methods: The
Non-Commutative Kubo Formula for the Hall conductivity}}). Periodic
boundary conditions are applied along the $x$- and $y$-directions,
while open boundary conditions are used along the $z$-direction to
model the finite thickness of a thin film. The calculated $\sigma_{xy}$
in the $E_{\rm F}-W$ plane is shown in Fig. \ref{fig:phase2}(a). A pronounced
Hall plateau, indicated by the bright yellow region, is observed in
the lower-left corner of the phase diagram. This plateau demonstrates
that the Hall conductivity remains robust under moderate disorder
even when the chemical potential deviates from the Dirac point, but
is still located with the gap of the top surface states induced by
magnetic doping. Adjacent to this region, a non-quantized anomalous
Hall effect, indicated by chartreuse region, is observed, and gradually
decays to vanish with increasing the disorder strength, as indicated
by the dark purple region. To investigate the behavior of $\sigma_{xy}$
in the thermodynamic limit, we perform a finite-size analysis of the
Hall conductivity (see Supplementary Note 1) along
a vertical line at a fixed disorder strength $W=1$ eV
as shown in Fig. \ref{fig:phase2}(c-1). The Fermi energy $E_{\rm F}$
is varied from 0 to $0.2$ eV, with the color gradient transitioning
from red to blue in increments of $0.02$ eV. For $E_{\rm F}<0.06$ eV,
where the Fermi level $E_{\rm F}$ intersects a single gapless Dirac cone,
$\sigma_{xy}$ quickly converges to $\frac{1}{2}\frac{e^{2}}{h}$ as the system size
$L$ varies from 12 to 28. In this region, the $\sigma_{xy}$ nearly
collapses onto a single curve (dashed line) and follows a power-law
behavior: 
\begin{equation}\begin{aligned}
\sigma_{xy}=\sigma_{xy}^{0}\left[1-\left(\frac{l_{e}}{L}\right)^{4}\right]
\end{aligned}\end{equation}
in units of $\frac{e^{2}}{h}$ , where $l_{e}$ is a characteristic
length, and $\sigma_{xy}^{0}$ denotes the Hall conductivity in the
thermodynamic limit. As exhibited in Fig. \ref{fig:phase2}(c-2),
the extracted value of $\sigma_{xy}^{0}$ is $0.49997\pm0.00021$,
closely approaching the half-integer quantization with high precision.
The scaling behavior of $\sigma_{xy}$ differs fundamentally from
that of the integer quantum Hall effect, which is strictly quantized
to integer values (0 or 1) \cite{Pruisken}. The shift of $\sigma_{xy}$
by $1/2$ from the conventional integer quantum Hall effect can be
interpreted as a manifestation of the parity anomaly, whose robustness
against disorder highlights the exotic and unique topological nature
of the HQHE. When $E_{\rm F}$ intersects a single gapless Dirac cone,
the random gapless Dirac Hamiltonian belongs to the symplectic symmetry
class with time reversal symmetry. However, the bottom gapless surface
states are connected to the top gapped surface states through the
side surface, effectively making the gapped surface states act as
a time-reversal symmetry-breaking boundary condition for the gapless
Dirac cone \cite{Ostrovsky2007:PRL,Hiroaki2016:JPSJ,Zou2022:PRB}.
As a consequence, the gapless Dirac cone on the surface of a semimagnetic
TI film preserves time-reversal symmetry in the bulk but breaks time-reversal
symmetry on the boundary. This results in nontrivial topology, distinguishing
it from the ordinary symplectic symmetry class. As shown in Ref. \cite{Zou2022:PRB},
a half-quantized Hall current circulates around the boundary between
the gapped and gapless Dirac cone regions. This corresponds to a half-quantized
Hall effect and reflects the parity anomaly of gapless Dirac fermions
in condensed matter systems. In the framework of effective field theory \cite{Ostrovsky2007:PRL},
implementing an infinite-mass boundary condition as a regularization
scheme introduces a topological $\theta$ term into the low energy
effective action of the nonlinear sigma model for the disordered single
gapless Dirac cone, with the topological angle fixed at $\theta=\pi$
(with the sign in our case depending on the sign of the Zeeman field,
$\mathrm{sgn}(V_{0})$). The topological term exerts minimal influence
in the metallic regime with a large longitudinal conductivity $\sigma_{xx}$,
but should crucially alter the renormalization group flow in the strong
disorder regime with small $\sigma$. For small $\sigma$, the theory
of a gapless single-node Dirac fermion is analogous to the critical
line of the integer quantum Hall transition. There exists an attractive
fixed point $\sigma^{*}$, such that as the temperature decreases
the system is driven into a metallic state exhibiting a half-quantized
Hall effect, characterized by $(\sigma_{xx},\sigma_{xy})=(\sigma^{*},\pm\frac{e^{2}}{2h})$.
This behavior may explain the temperature-dependent flow of longitudinal
and Hall conductance observed in semi-magnetic topological insulators
at zero magnetic field \cite{Mogi2022:NatPhys}. For large $\sigma$,
the system exhibits a metal-insulator transition at a critical point
$\sigma_{sp}$, similar to the behavior observed in the symplectic
class \cite{markovs2006critical,asada2004numerical}. Several numerical
investigations indicate that a single gapless Dirac cone with nonmagnetic
disorder potential is always metallic with a positive beta function
\cite{Bardarson:PRL,Nomura2007:PRL}. However, these studies are typically
conducted in momentum space by imposing a hard cutoff at a sufficiently
large momentum. This approach overlooks the influence of boundary
effects with symmetry breaking, which can play a significant role
in the system's transport behavior. Despite differing in certain aspects,
these studies consistently conclude that the isolated single gapless
Dirac cone state remains metallic in the presence of disorder when
the Fermi energy $E_{\rm F}$ is near the Dirac point.

When $E_{\rm F}$ is larger than $0.06$ eV, the Fermi surface intersects
both gapless and gapped surface Dirac cones. In the absence of disorder,
the Hall conductivity is non-quantized due to the breaking of time-reversal
symmetry in the two Fermi surface loops \cite{fu2025z}. When disorder
is introduced, the Hall conductivity still shows an increasing trend
with growing system size $L$ as shown in yellow and blue lines in
Fig. \ref{fig:phase2}(c-1). Our numerical analysis reveals that in
this regime $\sigma_{xy}$ does not seem to converge to a quantized
value. This observation contradicts the expected physical picture,
where disorder is expected to localize states associated with the
gapped Dirac cone, and the Hall conductivity should solely arise from
the delocalized gapless Dirac cone, scaling the system to $\frac{1}{2}\frac{e^{2}}{h}$.
However, this simplified picture overlooks the potential impact of
inter-surface scattering, which could significantly modify the underlying
physics. As $E_{\rm F}$ increases further toward the high-energy band
edge, the system transitions into an Anderson insulating state, where
$\sigma_{xy}$ scales to $0$. Our numerical results reveal that the
crossover region between the metallic phase with $\sigma_{xy}=\frac{1}{2}\frac{e^{2}}{h}$
and Anderson insulator state with $\sigma_{xy}=0$ actually constitutes
a marginal (or critical) phase exhibiting non-quantized Hall conductivity,
instead of shrinking to a single critical point. Similarly, a finite-size
analysis of $\sigma_{xy}$ along the horizontal axis at $E_{\rm F}=0.01$eV
is shown in Fig. \ref{fig:phase2}(d-1). Here, as $L$ increases,
$\sigma_{xy}$ exhibits a pronounced saturation to the half-quantized
value up to a critical disorder strength $W_{\rm c}\approx2.0$
eV. The determined $\sigma_{xy}^{0}$ value fluctuates within a narrow
margin around the half-integer plateau, as depicted in Fig. \ref{fig:phase2}(d-2),
with extracted bounds spanning between $0.4993\pm3.735\times10^{-4}$
and $0.5002\pm1.929\times10^{-4}$. Beyond this critical strength,
the Hall conductivity begins to deviate significantly from the quantized
value.

The lower-left corner of the phase diagram, corresponding to weak
disorder and small Fermi energy $E_{\rm F}$ represents the HQHM characterized
by a half-quantized Hall effect and metallic behavior. In 2D or quasi-2D
systems, the Hall effect of an insulating state must be integer-quantized,
and this result remains robust even in the presence of disorder due
to the topological nature of the quantum Hall effect. In the phase
diagram, regions outside the HQHM state that exhibit a non-zero but
non-quantized Hall conductivity must correspond to a metallic state.

To understand the origin of the transition from HQHE to non-quantized
AHE, we can employ the effective medium theory in conjunction with
the Kubo formula for electrical conductivities \cite{Bastin1971:JPCS,ChenYu2018:PRB}
(see \hyperref[subsec:Model]{\textit{Model and Material Realization}} and \hyperref[subsec:EMT]{\textit{Methods: Effective
Medium Theory}}), and the results are presented in \ref{fig:phase2}(b).
The self-consistent Born approximation is a powerful tool and applied
extensively to investigate the physics in topological Anderson insulator
\cite{Groth2009:PRL,Guo2010:PRL,Chen2015:PRL,Chen2017:PRB,Chen2018:PRB,Hua2019:PRB,Li2020:PRL,WangXR2020:PRR,Sarma2022:PRB,Chen2023:PRB}.
We extend the self-consistent Born approximation to layered structure.
After averaging over disorder, the Green's function recovers the translational
symmetry: $G^{R}=\langle[E_{\rm F}+\mathrm{i}0^{+}-H_{0}-H_{\mathrm{imp}}]^{-1}\rangle=\left(E_{\rm F}-H_{0}-\Sigma^{R}\right)^{-1}$.
Both the mass, magnetic gap, and the chemical potential are renormalized
by the real part of the self-energy, with $m_{0}\to\widetilde{m}_{0}$,
$V_{0}\to\widetilde{V}_{0}$ and $E_{\rm F}\to\widetilde{E}_{\rm F}$. As
depicted in Fig. \ref{fig:phase2}(b), the renormalized Dirac mass
$\widetilde{m}_{0}$ (orange $\square$ line) is remarkably enhanced
with increasing disorder strength, directly accounting for the persistent
gapless Dirac cone observed in Fig. \ref{fig:phase1}(b). The imaginary
part of the self-energy, $\eta$, is crucial for understanding the
non-quantization of $\sigma_{xy}$. $\eta(E_{\rm F})$ depends on the
energy $E_{\rm F}$ and determines the band broadening for each band.
In the absence of disorder, when chemical potential lies within magnetic
gap $|E_{\rm F}|<|V_{0}|$, half-quantized Hall currents propagate on
the gapless surface without dissipation \cite{Zou2022:PRB}. However,
disorder renormalizes the magnetic gap and abruptly triggers an energy
level broadening $\eta_{\mathrm{top}}$ (orange $\triangle$ line)
for the top surface states at a critical value $W_{\mathrm{c}}\approx2.6$
eV. Beyond $W_{\mathrm{c}}$, this broadening $\eta_{\mathrm{top}}$
becomes more pronounced and rapidly exceeds multiples of the magnetic
gap, as evidently demonstrated in Fig. \ref{fig:phase2}(b). The effective
coupling between two chiral currents, arising from the impurity scattering-induced
$\eta_{\mathrm{top}}$, leads to quantization errors and ultimately
completely destroying the conducting edge currents. Hence, when accounting
for disorder effects, the condition for quantization of Hall conductivity
should be modified as $|\widetilde{E}_{\rm F}|<|\widetilde{V}_{0}|-\eta_{\mathrm{top}}$.
Especially, for $E_{\rm F}=0$, when the disorder exceeds the threshold
value $W_{\mathrm{c}}$ (where $|\widetilde{V}_{0}|=\eta_{\mathrm{top}}(E_{\rm F}=0)$),
the band broadening induced by impurity scattering fills the magnetic
gap, causing the Hall conductivity to deviate from its quantized value
and ultimately decay to zero. We further compute $\sigma_{xy}$ by
integrating the Kubo-Bastin formula with the effective medium theory.
Below the threshold $W_{\mathrm{c}}$, $\eta_{\mathrm{top}}$ diminishes,
yielding a distinct half-quantized plateau in $\sigma_{xy}$ (blue
$\circ$ line), as illustrated in Fig. \ref{fig:phase2}(b). The black
dashed line marks the onset of $\eta_{\mathrm{top}}$. As $\eta_{\mathrm{top}}$
further increases, $\sigma_{xy}$ rapidly deviates from half-quantization
simultaneously, thereby substantiating our preceding analysis. Fig.
\ref{fig:phase2}(a) demonstrates the phase boundary (denoted by the white
circle line), derived from the effective medium theory, which aligns
quantitatively with numerical results. Ultimately, the effective medium
theory elucidates the mechanism underlying the breakdown of HQHE.

\paragraph*{Marginal Metal and Its Transition to Anderson Localization}\label{subsec:MM}

\begin{figure}[H]
\centering
\includegraphics[width=12cm]{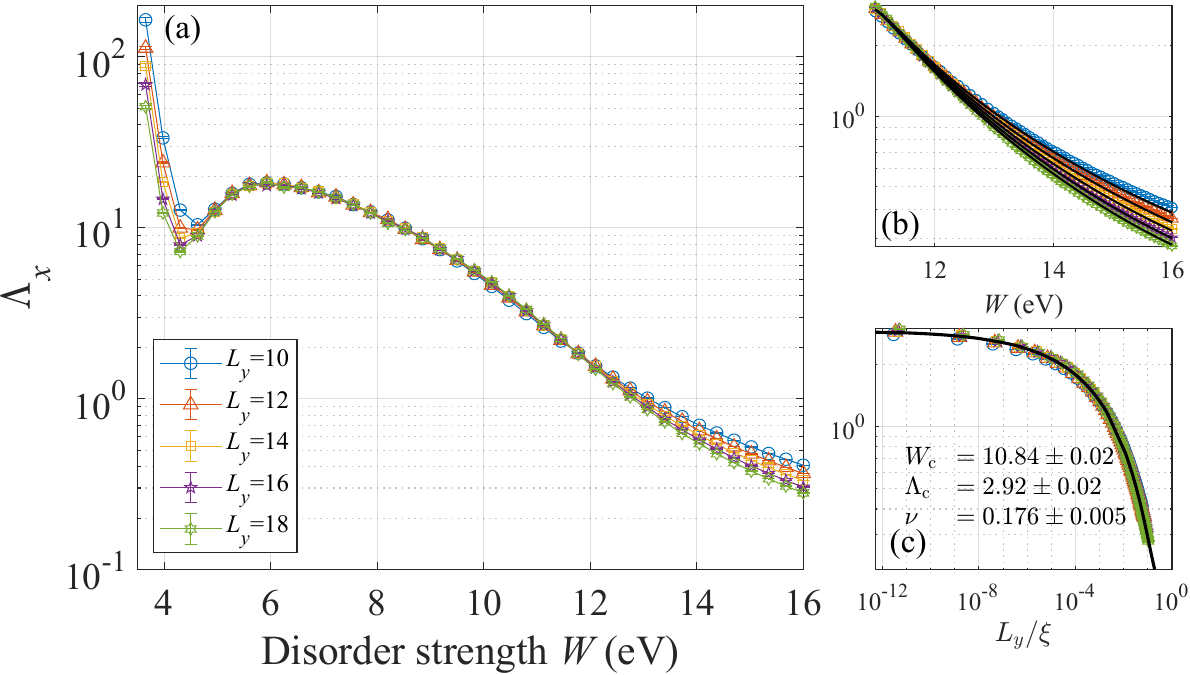}
\caption{\textbf{The normalized localization length and the emergence of the
marginal metal region in moderate disorder.} (a) The normalized localization
length $\Lambda_{x}$ along $x$ direction, where the error bars show standard deviations
of $\Lambda_{x}$ \cite{Yamakage2013:PRB}. The used longitudinal
size is $L_{x}=2\times10^{6}$. The thickness is $L_{z}=10$, and
we set Fermi energy at $E_{\rm F}=0.01$ eV. In the marginal metallic
phase, the normalized localization length exhibits a unique behavior
by collapsing onto a single line and showing no size dependence, demonstrating
its exotic criticality. (b) The normalized delocalization length near
the transition point. (c) The fit of the numerical data near the transition
point via a universal scaling function $\Lambda_{x}=F\left(L_{y}/\xi\right)$.
The transition points and the critical exponents are enumerated in
the figure.}
\label{localization-length}
\end{figure}

To explore the quantum criticality in semimagnetic TI induced by disorder,
we conduct a comprehensive investigation of localization physics by
employing the transfer matrix method (see \hyperref[subsec:NLL]{\textit{Methods: Normalized
Localization Length}}). The transfer matrix method is used to calculate
the localization length in a quasi-1D geometry, which consists of
a bar with a cross-section of $L_{y}\times L_{z}$ in the transverse
direction and infinite length $L_{x}$ in the longitudinal direction.
In the calculations, the thickness of the film is fixed at $L_{z}=10$
and the width $L_{y}$ is varied from 10 to 18. The normalized localization
length $\Lambda_{x}$ along $x$-direction encodes the information
of localization and delocalization. If $\Lambda_{x}$ grows with the
system size $L_{y}$, then in the thermodynamic limit $L_{y}\to+\infty$,
$\Lambda_{x}$ becomes divergent, indicative of a metallic state.
Conversely, it would become vanishing and describes a localized state.
At the quantum critical point, $\Lambda_{x}$ does not change with
$L_{y}$ and exhibits scale-invariant behavior as a marginal or critical
metal. As depicted in Fig. \ref{localization-length}, our analysis
reveals that, for extremely strong disorder, the normalized localization
length exhibits a decreasing trend with respect to system width $L_{y}$,
thereby providing a definitive manifestation of Anderson localization.
The unexpected discovery elicited significant surprise when it was
observed that within the moderate disorder regime prior to the onset
of the Anderson localization, the normalized localization length converges
onto a single line, defying expectations of an isolated critical point,
thus underscoring the unconventional scale-invariant behavior of the
system in this regime. This finite scale-invariant region suggests
the emergence of a marginal metal \cite{chen2019metal} (also named
critical metal \cite{wang2024anderson}) phase rather than a single
critical point, delineating a unique and unconventional form of quantum
criticality reminiscent of the Berezinskii-Kosterlitz-Thouless (BKT)
phase transition \cite{xie1998kosterlitz,ChenCZ2019:PRL,wang2015band,zhang2009localization}.
Further reducing the disorder strength (approximately below 5 eV),
the normalized localization lengths increase drastically up to one
or two orders, which far exceeds the system sizes of the cross section
and might indicate the potential limitations or unreliability of the
approach. Instead we shall use an alternative technique to tackle
the problem in the weak disorder regime.

In order to delve deeper into the enigmatic disorder-induced quantum
phase transition, we adhere to the traditional one-parameter scaling
hypothesis and conduct finite-size scaling analysis in the vicinity
of the transition point. The data for the normalized localization
length in proximity to the transition point conform to a universal
function $\Lambda_{x}=F\left(L_{y}/\xi\right)$, which can be expressed
through a Taylor series expansion up to the second order
\begin{equation}\begin{aligned}
\Lambda_{x}=\Lambda_{\mathrm{c}}+c_{1}\left(\frac{L_{y}}{\xi}\right)^{\nu}+c_{2}\left(\frac{L_{y}}{\xi}\right)^{2\nu},
\end{aligned}\end{equation}
where $\xi$ represents the correlation length and exhibits a divergent
behavior within the quantum critical phase. Consequently, we interpret
$\xi$ as embodying the form $\xi=\exp\left[b/\sqrt{\frac{W-W_{\mathrm{c}}}{W_{\mathrm{c}}}}\right]$,
which is highly evocative of the BKT criticality arising in the 2D
$XY$ spin model \cite{Kosterlitz1974:JPC}. $b$ is a non-universal
number that depends on the system parameters. Numerical
investigations of BKT criticality in disorder-driven quantum phase
transitions have been carried out in various systems, including 2D
electron gas with random magnetic flux \cite{xie1998kosterlitz},
Graphene with long-range impurities \cite{zhang2009localization},
$C_{4z}\mathcal{T}$ network model \cite{wang2024anderson}, 2D random
SU(2) electron gas in a constant magnetic field \cite{wang2015band},
2D Rashba electron gas with a Zeeman split \cite{chen2019metal},
and spin Chern insulator with structure inversion asymmetric impurity
potential \cite{ChenCZ2019:PRL}, among others. Our finite-size scaling
analysis yields the exponent $\nu=0.176\pm0.005$
for the MM-to-AI transition as shown in Fig. \ref{localization-length},
which raises an open question regarding its universality. Further
investigation into the application of the renormalization group calculations
would be merited in order to systematically derive the 
exponents, thereby shedding light on the underlying mechanism governing
this phenomenon.

\paragraph*{Quantum Interference Effect and Phase Transition}\label{subsec:QIE}

We now investigate the metallicity of the system through the lens
of longitudinal conductivity $\sigma_{xx}$. Unlike the Hall conductivity,
$\sigma_{xx}$ is primarily determined by the scattering of electrons
near the Fermi surface, making it a sensitive probe of localization
behavior at $E_{\rm F}$. The Ioffe-Regel criterion, $k_{\rm F} l \lesssim 1$,
defines a strong scattering regime where the nature of wave transport
remains ambiguous \cite{ioffe1960non}. Here $k_{\rm F}=E_{\rm F}/\hbar\lambda_{\parallel}$
is the Fermi wave number, and $l$ is the mean free path. The condition
$k_{\rm F}l\sim1$ is often associated with the onset of Anderson localization,
while $k_{\rm F}l>1$ corresponds to a weak scattering regime. In the
latter, quantum interference between time-reversal symmetric counter-propagating
paths can lead to quantum corrections in $\sigma_{xx}$, resulting
in weak localization (WL) or WAL, depending on the system's symmetry
\cite{HLN1980:PTP,lu2011competition,lu2013extrinsic}.

To explore the quantum transport properties, we employ the quantum
interference theory to evaluate the lowest-order quantum correction
$\delta\sigma_{qi}$ to the conductivity, which is valid in the weak
scattering regime (see Supplementary Note 4 for
detailed calculations).

\begin{figure}[ht!]
\centering
\includegraphics[width=12cm]{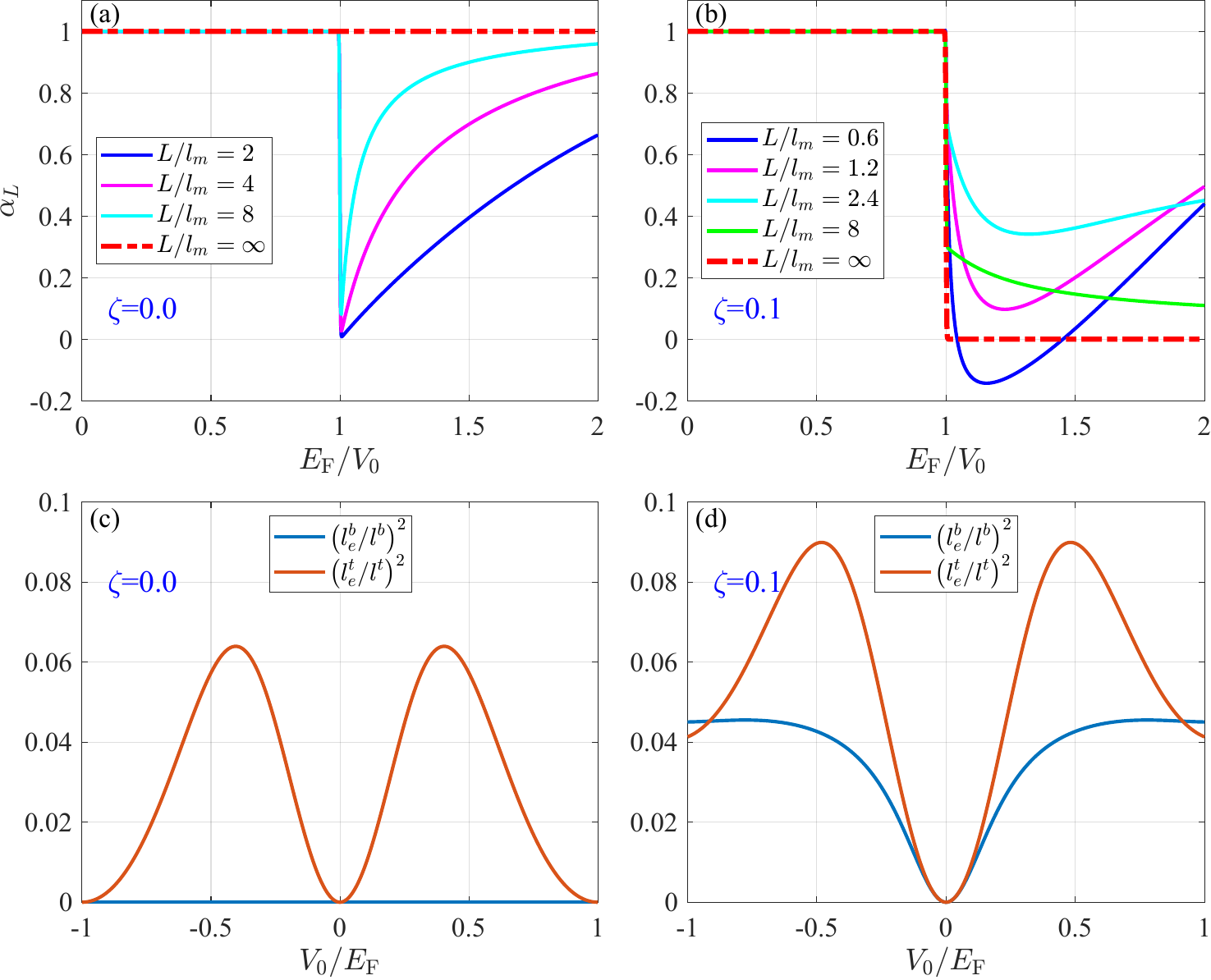}
\caption{\textbf{Results of quantum interference corrections.} Dependence of
$\alpha_{L}=\frac{\pi h}{e^{2}}\frac{\partial\sigma}{\partial\ln L}$
on $E_{\rm F}/V_{0}$ for different inter-surface scattering strength
(a) $\zeta=0$ and (b) $\zeta=0.1$ with $L/l_{m}$ varied and $l_{m}=\frac{\lambda_{\parallel}^{2}}{V_{0}U_{0}}$. The characteristic length scales $(l_{e}^{s}/l^{s})^{2}$ for two
Cooperon channels ($s=t,b$) as a function of $V_{0}/E_{\rm F}$ for (c)
$\zeta=0$ and (d) $\zeta=0.1$. Here, $l_{e}^{s}$ represents the
mean free paths for the respective channels.}
\label{fig:lcond}
\end{figure}

(i) When $E_{\rm F}$ lies within the magnetic gap of the top (t) surface
states and only intersects the bottom (b) surface states ($E_{\rm F}<V_{0}$),
the problem simplifies to the localization of a single gapless Dirac
Hamiltonian. The quantum correction is found to scale with the system
size $L$ as: $\delta\sigma_{qi}\simeq\frac{e^{2}}{\pi h}\ln(L/l_{e})$.
The degeneracy point in momentum space acts as a Dirac monopole, causing
the wave functions to acquire a $\pi$ phase shift when transported
around the Fermi surface in the Berry connection. This results in
destructive quantum interference, leading to WAL. In the WAL regime,
$\sigma_{xx}$ exhibits a logarithmic divergence with system size
$L$ with $\alpha_{L}=+1$ (Fig. \ref{fig:lcond}(a)
and (b)).

(ii) When $E_{\rm F}$ lies outside the magnetic gap, it intersects two
bands of the surface states. As shown in Supplementary Note 4,
the full expression for $\alpha_{L}$ can be expressed as a sum over
two Cooperon channels: $\alpha_{L}=\sum_{s=t,b}\frac{\alpha^{s}}{(L/l^{s})^{2}+1}$
where $\alpha^{s}$ and $l^{s}$ are the weighting factors and characteristic
lengths associated with the two channels, respectively. Here, $l^{s}$
measures the suppression of quantum interference effects due to dephasing
or symmetry-breaking mechanisms. It determines the decay length of
the Cooperon mode: a smaller $l^{s}$ corresponds to stronger suppression
of quantum interference, while a larger $l^{s}$ allows for more pronounced
interference effects. In the absence of inter-surface scattering,
the two surface states are independent, and the total quantum correction
is simply the sum of the individual contributions from the two surface
states. The bottom surface exhibits genuine WAL with $\alpha^{b}=1$
and $l^{t}\to\infty$, shown as the red line in Fig. \ref{fig:lcond}(c).
In contrast, the top surface states show a crossover from WL ($\alpha^{t}=-1,l^{t}\to\infty$)
to WAL ($\alpha^{t}=+1,l^{t}\to\infty$) as $E_{\rm F}$ varies from band
edge ($E_{\rm F}\sim V_{0}$) to high energy $(|E_{\rm F}|\gg|V_{0}|)$. This
occurs because, in these two limiting regimes, an orthogonal type
and a symplectic type of time-reversal symmetry are recovered on the
Fermi surface, even though the magnetic gap explicitly breaks time-reversal
symmetry. In other cases, the bottom surface channel exhibits suppressed
WL or WAL with finite $l^{t}$. As $L$ increases, the contribution
from the top surface channel is suppressed following $\sim1/[(L/l^{t})^{2}+1]$.
In the thermodynamic limit, only the top surface channel survives,
and $\alpha_{L}$ approaches $+1$ (except at $E_{\rm F}/V_{0}=1$). In
the presence of inter-surface scattering, the scaling behavior of
$\alpha_{L}$ becomes qualitatively different. Inter-surface scattering
introduces a finite Cooperon gap to the bottom surface's Cooperon
channel, resulting in a finite $l^{b}$, as shown in Fig. \ref{fig:lcond}(d).
Consequently, the contribution from the bottom channel is also suppressed,
following $\sim1/[(L/l^{b})^{2}+1]$. In this case, for $E_{\rm F}/V_{0}>1$,
$\alpha_{L}$ approaches $\sum_{i=t,b}\alpha^{i}/[(L/l^{i})^{2}+1]\rightarrow0$
instead of $1$ in the limit of large $L$, as illustrated by the
dashed line. We conclude that this crossover behavior in the scaling
of $\alpha_{L}$ around $V_{0}$ arises due to inter-surface scattering.

When $E_{\rm F}<|V_{0}|$, the $\pi$ Berry phase, inherent to a single
gapless Dirac cone, serves as the foundation for the interconnected
phenomena of WAL and the half-quantized Hall effect. Arising from
the Dirac cone's topological structure, the $\pi$ Berry phase drives
WAL by enhancing conductivity through quantum interference in the
presence of spin-orbit coupling \cite{lu2011competition,lu2013extrinsic}.
Simultaneously, the anomalous Hall conductivity in metallic systems
can be linked to the Berry phase accumulated over the Fermi surface
loop, enabling the half-quantized Hall effect under broken time-reversal
symmetry \cite{fu2025z}. Together, these concepts underscore the
profound relationship between topology, quantum interference, and
electronic transport in topological metallic systems, setting them
apart from the integer quantum Hall effect which is exclusively associated
with insulating phases. When $E_{\rm F}$ intersects two surface states
bands, we confirm that inter-surface scattering events introduce finite
Cooperon gaps across all channels, leading to a strong suppression
of quantum interference corrections. To conclusively determine the
phase in the thermodynamic limit, high-loop quantum correction calculations
are necessary. Recent numerical studies have shown that the MM can
emerge in two-dimensional ferromagnetic electron gases with random
scalar potentials and Rashba spin-orbit interactions, where time-reversal
symmetry breaking, spin-orbit coupling, and interband scattering are
essential ingredients \cite{chen2019metal}. In contrast, we demonstrate
a distinct scenario where the marginal metal phase is mediated by
residual Berry curvature and weak inter-surface scattering, highlighting
a different mechanism for realizing this phase in disordered semimagnetic
topological insulators. For a finite system of size $L$, the correlation
length $\xi$ is bounded by $L$. Near a critical point, temperature
$T$ and sample size $L$ can be related through scaling arguments:
$\rm{d}\ln L/\rm{d}\ln T=-\nu_{T}$ with $\nu_{T}$ as the critical exponent
which reflects how the temperature scales relative to the sample size.
In MM phase, where $\alpha=0$, the longitudinal conductivity $\sigma_{xx}$
becomes temperature-independent. This distinctive feature can serve
as a key experimental signature for identifying the MM phase. Finally,
we emphasize that while reducing the film thickness can enhance inter-surface
scattering, thereby making the observation of the MM phase in ultrathin
films more promising, the HQHM phase tends to be suppressed, as the
surface Dirac cones are more likely to develop a finite-size energy
gap \cite{lu2010massive}.

\paragraph*{The Arithmetic and Geometric Mean Density of States}\label{subsec:TwoDOS}

The emergence of the metallic phases in semimagnetic TI can be characterized
by analyzing the two types of means of the local density of states
(DOS): the arithmetic mean $\rho_{\mathrm{a}}$ and the geometric
mean $\rho_{\mathrm{t}}$ (see \hyperref[subsec:dos]{\textit{Methods: Density of States}}).
The local DOS, $\rho_{{\bf r},\alpha}(E)=\left\langle {\bf r},\alpha\right|\delta\left(E-H\right)\left|{\bf r},\alpha\right\rangle $,
quantifies the amplitude of the wave function at site ${\bf r}$ for
a given energy $E$, where $\left|{\bf r},\alpha\right\rangle $ denotes
an $\alpha$-orbital electron wave function at that site. The spatial
distribution of $\rho_{{\bf r},\alpha}(E)$ contains direct information
about the localization properties, which are closely intertwined with
the topology of the quantum system \cite{Pruisken,Tian2016:PRB,Zhang1994:PRL}.
The arithmetic and geometric mean DOS for a disordered system are
defined as $\rho_{\mathrm{a}}(E)=\left\langle \frac{1}{V}\sum_{i=1}^{V}\sum_{\alpha=1}^{4}\rho_{{\bf r}_{i},\alpha}(E)\right\rangle _{\mathrm{imp}}$
and $\rho_{\mathrm{t}}(E)=\exp\left[\frac{1}{N_{s}}\sum_{i=1}^{N_{s}}\left\langle \ln\sum_{\alpha=1}^{4}\rho_{{\bf r}_{i},\alpha}(E)\right\rangle _{\mathrm{imp}}\right]$,
respectively \cite{ZhangYY2012:PRB,ZhangYY2013:PRB}. A zero value
of arithmetic mean DOS $\rho_{{\rm a}}(E)$ at a finite interval defines
an energy gap, such as a band insulator, while a nonzero value of
$\rho_{{\rm a}}(E)$ means that there exists the states at the energy
$E$, which can be either localized or delocalized. The geometric
mean DOS $\rho_{{\rm t}}(E)$ may reveal more information for the
localization of the states. For energy $E$ where all states are extended,
local DOS are uniformly distributed throughout the system, and we
have $\rho_{{\rm t}}\simeq\rho_{{\rm a}}$. However, for the states
whose wave-functions are localized in real space, there is no contribution
to $\rho_{{\rm t}}$ \cite{KPM2006:RMP}. Generally, comparison of
$\rho_{\mathrm{a}}(E)$ to $\rho_{\mathrm{t}}(E)$ can investigate
the localization properties of particle states. In the case of quasi-2D
topological insulator films featuring surface states, the analysis
of these properties becomes more complex (See Supplementary Note 2
for detailed analysis). Further study shows that
since the surface states are localized near the top or bottom surface
in $z$ direction and extended in $x$ and $y$ direction, these states
do not contribute to the value of $\rho_{{\rm t}}$, but do to $\rho_{{\rm a}}$.
For a non-vanishing $\rho_{{\rm a}}$, if the ratio $\rho_{\mathrm{t}}/\rho_{\mathrm{a}}$
is 0, it can define two different phases: one is that all states are
localized in real space in all directions, i.e., AI phase, and another
one is that the states are only localized at the top or bottom surface,
i.e., the surface states. For a finite ratio of $\rho_{\mathrm{t}}/\rho_{\mathrm{a}}$
it indicates the existence of the delocalized states, defining a metallic
phase.

\begin{figure}[H]
\centering
\includegraphics[width=12cm]{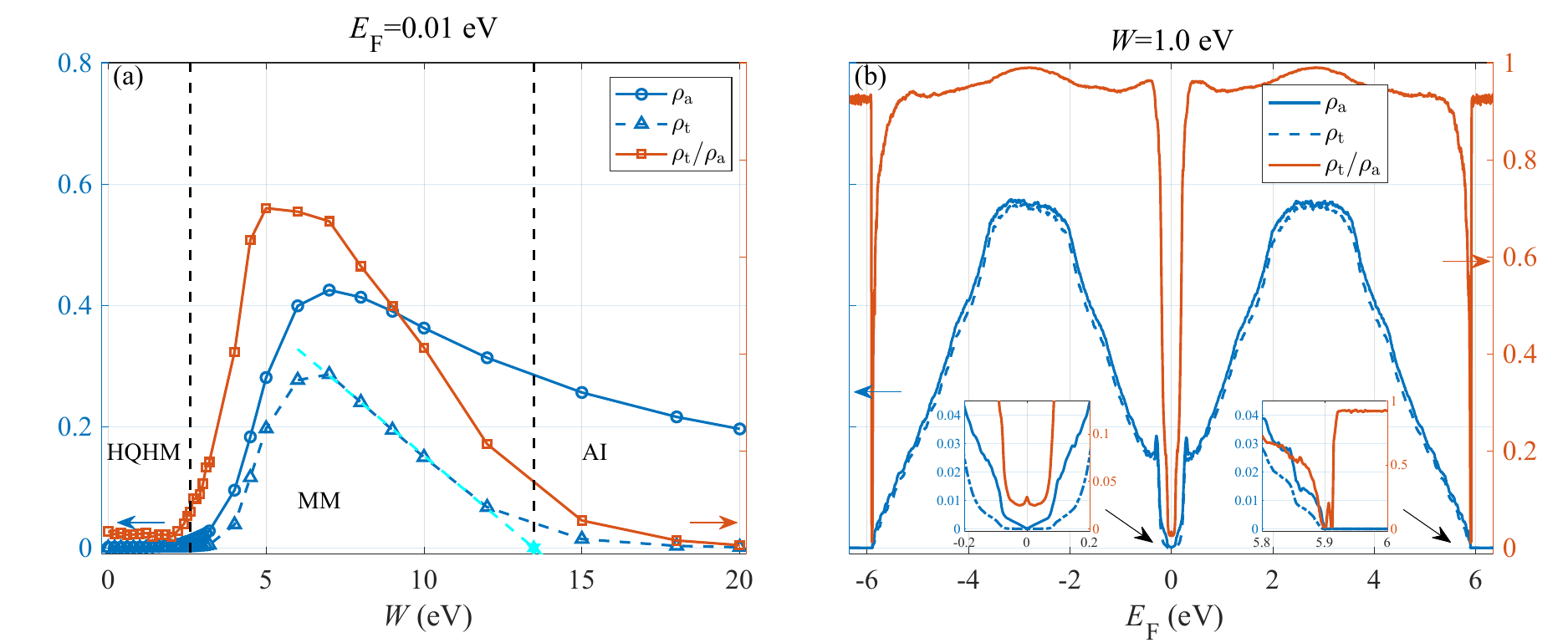}
\caption{\textbf{The arithmetic and geometric mean DOS} $\rho_{\mathrm{a}}(E)$
\textbf{and} $\rho_{\mathrm{t}}(E)$. (a) The DOS as the function
of disorder strength $W$ at the Fermi energy $E_{\rm F}=0.01$ eV, revealing the 
phase transitions between half-quantized Hall metal (HQHM), marginal metal (MM), and
Anderson insulator (AI). The cyan dashed line fitted to $\rho_{\mathrm{t}}$ intersects the horizontal
axis at the hexagonal marker ($W=13.47$ eV), thereby demarcating
the phase boundary between MM and AI. (b) The DOS as the function
of the Fermi energy $E_{\rm F}$ at $W=1.0$ eV. The insets show an 
enlarged view of the band center and band edge.}
\label{DOS}
\end{figure}

Fig. \ref{DOS} exhibits the arithmetic and geometric DOS as a function
of disorder at a fixed chemical potential and as a function of the
chemical potential at a fixed disorder, respectively. In Fig. \ref{DOS}(a),
we set $E_{\rm F}=0.01$eV, which is close to the Dirac point. The vanishing
$\rho_{\mathrm{a}}(E)$ to $\rho_{\mathrm{t}}(E)$ at weak disorder
indicates that the surface states are localized near one surface in
HQHM. The disorder-induced finite DOS appears in a moderate disorder,
serving as critical evidence for the MM phase. Both $\rho_{\mathrm{a}}(E)$
and $\rho_{\mathrm{t}}(E)$ tend to vanish for a strong disorder,
demonstrating the localization of electrons. Specifically,
we find that $\rho_{\mathrm{t}}(E)$ exhibits a nearly-linearly decay
behavior when approaching the MM-AI phase boundary. Therefore, we
linearly fit the $\rho_{\mathrm{t}}(E)$ data for $W$ from 7 eV to
12 eV and extrapolate to the intersection with the horizontal axis
at $W\approx13.47$ eV. We expect that in the thermodynamic limit,
$\rho_{\mathrm{t}}(E)$ is finite for MM phase, and vanishes for AI
phase. In Fig. \ref{DOS}(b), we take a weak disorder $W=1.0$ eV.
The geometric mean DOS $\rho_{\mathrm{t}}(E)$ is almost zero for
a chemical potential residing around the Dirac point ($E<0.06$ eV),
consistent with that in Fig. \ref{DOS}(a). The DOS $\rho_{\mathrm{t}}(E)$
becomes finite when the chemical potential intersects with the bands
of the massive Dirac fermions, indicating the metallicity of the system
even in the presence of disorder in this quasi-2D system.
The DOS $\rho_{\mathrm{t}}(E)$ approaches zero again when the Fermi
level is close to the top of the band (about 5.91 eV), indicating
the localization of the electrons. The calculated DOSs provide alternative
and substantial evidence to support the phase diagram in Fig. \ref{fig:phase1}.
Some key features are summarized in Tab. \ref{tab:summary}.

\begin{table}[htbp]
\centering
\begin{tabular}{|c|c|c|c|}
\hline 
phase & HQHM & MM & AI \tabularnewline
\hline 
\hline 
$W$ range & < 2.6 eV & 2.6 $\sim$ 13.47 eV & > 13.47 eV \tabularnewline
\hline 
$\alpha$ & 1 & 0 & -\tabularnewline
\hline 
$\sigma_{xy}$ & $\frac{1}{2}\frac{e^{2}}{h}$ & non-universal & 0\tabularnewline
\hline 
$\rho_{\mathrm{a}}$ & finite & finite & finite \tabularnewline
\hline 
$\rho_{\mathrm{t}}$ & 0 & finite & 0 \tabularnewline
\hline 
\end{tabular}

\caption{\textbf{Summary of the key characteristics of the distinct phases in semimagnetic
topological insulator films.} All data are taken at $E_{\rm F}=0.01$ eV.
The ``-'' symbol indicates that $\alpha$ is not applicable in the
regime of strong disorder. }
\label{tab:summary}
\end{table}

\paragraph*{Model and Material Realization}\label{subsec:Model}

We adopt the tight-binding model on a cubic lattice for the semi-magnetic
TI topological insulator, 
\begin{equation}
H_{0}=\sum_{{\bf r}_{i}}\Psi_{{\bf r}_{i}}^{\dagger}(M_{0}+V_{z})\Psi_{{\bf r}_{i}}+\sum_{{\bf r}_{i},\alpha=x,y,z}(\Psi_{{\bf r}_{i}}^{\dagger}\mathcal{T}_{\alpha}\Psi_{{\bf r}_{i}+{\bf e}_{\alpha}}+\mathrm{H.c.}), 
\end{equation}
where $\mathcal{T_{\alpha}}=t_{\alpha}\tau_{z}\sigma_{0}-\frac{\mathrm{i}\lambda_{\alpha}}{2}\tau_{x}\sigma_{\alpha}$
and $M_{0}=\left(m_{0}-4t_{\parallel}-2t_{z}\right)\tau_{z}\sigma_{0}$
with the lattice size $L_{x}\times L_{y}\times L_{z}$ \cite{Zhang2009:NatPhys,Chu2011:PRB,Shen2012:TI}.
$\Psi_{{\bf r}_{i}}^{\dagger}$ and $\Psi_{{\bf r}_{i}}$ are creation
and annihilation operators of an electron at site $i$. $\tau_{\alpha}$
and $\sigma_{\alpha}$'s are Pauli matrices acting on the orbital
and spin spaces, respectively. The magnetic doping is modeled by introducing
a Zeeman potential $V_{z}=V_{z}(i_{z})\tau_{0}\sigma_{z}$. $V_{z}(i_{z})=V_{0}$
for the top layers $i_{z}\leqslant L_{z}^{\mathrm{Mag}}$, and 0 otherwise.
The model in Eq. (1) incorporating four orbitals, $|P1_{-}^{+},\pm\frac{1}{2}\rangle$
and $|P2_{+}^{-},\pm\frac{1}{2}\rangle$, was proposed to describe
the topological nature of strong topological insulator $\mathrm{Bi_{2}Se_{3}}$
and $\mathrm{Bi_{2}Te_{3}}$ by taking the bulk band gap $m_{0}=0.28$eV.
The model has been extensively used to explore the physics of strong
topological materials. Finally, we follow the common practice in the
study of Anderson localization and introduce the Anderson disorder
$H_{\mathrm{imp}}=\sum_{\mathbf{r}_{i}}\Psi_{{\bf r}_{i}}^{\dagger}u_{{\bf r}_{i}}\tau_{0}\sigma_{0}\Psi_{{\bf r}_{i}}$
through random on-site energy $u_{{\bf r}_{i}}$ that is uniformly
distributed in $\left[-W/2,+W/2\right]$.

The semimagnetic TI has already been realized experimentally in Cr-doped
$\mathrm{(Bi,Sb)_{2}Te_{3}}$, in which the half-quantized Hall conductance
was measured \cite{Mogi2022:NatPhys}.

\section*{Discussions}\label{sec:Discussions}

Our primary analysis focuses on short-ranged non-magnetic
impurities, and we have also examined the robustness of the HQHE against
other disorder types. The robustness of the HQHE is intrinsically
linked to the protection of the gapless Dirac cone in the system.
For magnetic impurities, the impact depends critically on the spin
symmetry of the disorder: (i) Spin-flip in-plane magnetic disorders
(e. g., $\sigma_{x}$ and $\sigma_{y}$-type) effectively shift the
Dirac cone position while preserving its gapless nature, allowing
the HQHE to remain robust. (ii) In contrast, $\sigma_{z}$-type out-of-plane
magnetic disorder opens a gap in the Dirac cone, introducing fluctuations
in the Hall response. However, since the energy gap induced by $\sigma_{z}$-type
impurities vanishes under disorder averaging, the disorder-averaged
Hall conductivity still maintains its half-quantized value \cite{Yi2025:JPC}.
This demonstrates that while local fluctuations may occur, the topological
protection of the HQHE survives when disorder preserves the global-averaged
time-reversal symmetry. These results highlight the remarkable stability
of the HQHE against various disorder types, provided they maintain
the essential gapless nature of the Dirac spectrum either locally
or on average.

We employ a combination of numerical and analytical methods to determine
the phase diagram of a disordered semimagnetic TI, each with its own
merits and limitations. (i) The calculation of the Hall conductivity,
combined with effective medium theory, establishes the full phase boundary
between the HQHM and MM phase in the $(W,E_{\rm F})$ plane. These phases
exhibit distinct size-scaling behaviors: the HQHE scales to $\frac{1}{2}\frac{e^{2}}{h}$ in
the large system size limit ($L\rightarrow\infty$), while the MM
phase converges to a finite value below $\frac{1}{2}\frac{e^{2}}{h}$. Effective medium theory
further supports this transition, showing that the Hall conductivity
deviates from $\frac{1}{2}\frac{e^{2}}{h}$ as the system enters the MM phase due to the emergence
of an imaginary part in the self-energy. It is also further supported
by the calculation of the DOS. (ii) Quantum interference calculations,
based on a quasiparticle picture, are valid for weak disorder and
applicable to few-band systems. They confirm the phase transition
between the HQHE metallic phase, where $E_{\rm F}$ intersects a single
gapless Dirac cone, and the MM phase, where $E_{\rm F}$ intersects two
surface state bands at weak disorder. The transition point is consistent
with that of the Hall conductivity. (iii) Localization length calculations
using the transfer matrix method are particularly effective in strongly
disordered regimes, where the localization length is small. This method
accurately captures the transition from the scale-invariant MM phase
to the Anderson insulator phase. (iv) To probe the localization nature
of bulk and surface states, we compute the arithmetic mean $(\rho_{\mathrm{a}})$
and geometric mean $(\rho_{\mathrm{t}})$ DOS. For weak disorder,
surface states localized on the surface of the quasi-2D film exhibit
a vanishingly small $\rho_{\mathrm{t}}$, distinct from conventional
extended states in purely 2D systems. In the MM phase, bulk-extended
states lead to a significant increase in $\rho_{\mathrm{t}}$. By
integrating all these results, we believe that a comprehensive and
reliable phase diagram has been established for the disordered semimagnetic
topological insulator.

In purely 2D systems, Anderson localization depends critically on
the symmetry class of the system. These symmetries are broadly categorized
into three main classes: orthogonal, unitary, and symplectic \cite{Mirlin2008:RMP}.
In the orthogonal and unitary symmetry classes, all states in 2D are
localized for any amount of disorder \cite{GoF1979:PRL,50YoAL}. However,
in the symplectic class, which preserves time-reversal symmetry and
includes strong spin-orbit coupling, the system exhibits a metal-insulator
transition as a function of disorder strength \cite{asada2002anderson,asada2004numerical}.
This transition arises due to weak anti-localization effects, which
counteract localization and allow for extended states at critical
disorder strengths. The integer quantum Hall transition further enriches
our understanding of localization by introducing the interplay between
Landau level quantization, disorder, and topology \cite{huckestein1995scaling}.
In the integer quantum Hall regime, extended states emerge at critical
energies within Landau levels, while all other states remain localized.
This transition is characterized by a change in the topological invariant
(Chern number) and is fundamentally distinct from conventional Anderson
localization. The integer quantum Hall transition exemplifies how
topology can stabilize extended states even in the presence of disorder.
The behavior of semimagnetic TI films, however, presents a unique
and complex scenario that defies simple classification. At first glance,
one might assume that magnetic doping on the surface breaks the global
time-reversal symmetry, placing the system in the unitary symmetry
class, where all states would be localized for any amount of disorder.
However, this is not the case. The quasi-2D nature of TI films, combined
with their nontrivial topological properties, leads to distinct features
that differentiate them from purely 2D systems. The surface state
on the undoped side of the film retains higher symmetry (symplectic
time-reversal symmetry), protecting it from complete localization
in weak disorder. For high-energy states that extend into the bulk
and interact with the magnetic surface, the symmetry is effectively
lowered, causing these states to become localized. In the intermediate
energy regime, instead of a single critical point, there exists a
finite region where critical extended states can emerge. As a result,
the localization behavior in semimagnetic topological insulator films
depends strongly on the position of the chemical potential. Different
energy regimes exhibit distinct local symmetries, leading to varied
localization behaviors. The interplay between magnetic doping, topological
protection, and spatial symmetry breaking leads to a rich variety
of localization behaviors, distinct from those in conventional 2D systems.

\section*{Methods}\label{sec:Methods}

\paragraph*{The Non-Commutative Kubo Formula for the Hall conductivity}\label{subsec:RSCN}

The Hall conductivity of a disordered system on a lattice can be calculated
using the non-commutative Kubo formula \cite{Prodan2011:JPA}
\begin{equation}
\sigma_{xy}=\frac{e^{2}}{h}\left\langle 2\pi\mathrm{i}\mathrm{Tr}\left\{ P\left[-\mathrm{i}\left[x,P\right],-\mathrm{i}\left[y,P\right]\right]\right\} \right\rangle _{\mathrm{imp}},
\label{eq:hall-conductivity}
\end{equation}
where $P$ represents the projector onto the occupied states, and
$x$ and $y$ are the coordinate operators on the lattice, respectively.
$\left\langle \cdots\right\rangle _{\mathrm{imp}}$ denotes the average
over the disorder sampling. We apply the periodic boundary condition
in both the $x$ and $y$ directions in order to minimize the boundary
effect while adopting open boundary condition in the $z$ direction
for a thin film.

\paragraph*{Density of States}\label{subsec:dos}

The spectral function is defined as 
\begin{equation}\begin{aligned}
A\left(\epsilon,{\bf k}\right)=\sum_{n}\left\langle n{\bf k}\right|\delta\left(\epsilon-H\right)\left|n{\bf k}\right\rangle,
\end{aligned}\end{equation}
where $H=H_{0}+H_{\mathrm{imp}}$ is the total Hamiltonian including
the impurity potential $H_{\mathrm{imp}}$ in real space. $\left|n{\bf k}\right\rangle $
is the eigenstate of the Hamiltonian $H_{0}$. To overcome the computational
complexity for a large system, we implement the Chebyshev polynomial
expansion of $A\left(\epsilon,{\bf k}\right)$ employing the kernel
polynomial method \cite{KPM2006:RMP,Fan2021:PhysRep}:
\begin{equation}\begin{aligned}
A\left(\epsilon,{\bf k}\right)=\frac{1}{\pi\epsilon_{\mathrm{max}}\sqrt{1-\widetilde{\epsilon}^{2}}}\sum_{n}\sum_{m=0}^{M-1}\left(2-\delta_{m,0}\right)g_{m}^{\mathrm{J}}T_{m}\left(\widetilde{\epsilon}\right)\left\langle n{\bf k}\right|T_{m}\left(\widetilde{H}\right)\left|n{\bf k}\right\rangle ,
\end{aligned}\end{equation}
in which $\epsilon_{\mathrm{max}}$ is chosen as an adequately large
energy scale ensuring that $\widetilde{\epsilon}=\epsilon/\epsilon_{\mathrm{max}}$
and eigenvalues of $\widetilde{H}=H/\epsilon_{\mathrm{max}}$ lie
within the domain of Chebyshev polynomials of the first kind $T_{m}\left(x\right)=\cos\left(m\arccos x\right)$. The Jackson damping kernel 
\begin{equation}\begin{aligned}
g_{m}^{\mathrm{J}}=\left(1-\frac{m}{M+1}\right)\cos\frac{m\pi}{M+1}+\frac{1}{M+1}\sin\frac{m\pi}{M+1}\cot\frac{\pi}{M+1}
\end{aligned}\end{equation}
serves to mitigate the Gibbs oscillation arising from truncating
the first $M$ term. Besides, the Jackson damping kernel has been
identified as the optimal choice for expanding the quantum resolution
operator, which is essentially a series of Dirac $\delta$ peaks,
as it minimizes broadening for a given value of $M$. More specifically,
away from $\widetilde{\epsilon}=\pm1$ it approximates the Dirac $\delta$
distribution with a Gaussian distribution with a width of $\delta\epsilon=\pi\epsilon_{\mathrm{max}}/M$.
In this work, we take a large lattice size $L_{x}=L_{y}=400$, and
$M=6000$ to achieve high resolution. Other parameters are the same
as those in Fig. \ref{fig:phase2}.

The localization properties can be characterized by analyzing the
two types of means of the local density of states (DOS): the arithmetic
mean $\rho_{\mathrm{a}}$ and the geometric mean $\rho_{\mathrm{t}}$.
The local DOS, $\rho_{{\bf r},\alpha}(\epsilon)=\left\langle {\bf r},\alpha\right|\delta\left(\epsilon-H\right)\left|{\bf r},\alpha\right\rangle $,
quantifies the amplitude of the wave function at site ${\bf r}$ for
a given energy $\epsilon$, where $\left|{\bf r},\alpha\right\rangle $
denotes an $\alpha$-orbital electron wave function at that site.
We derive the local DOS $\rho_{{\bf r},\alpha}(\epsilon)$ via substitution
into the aforementioned Chebyshev polynomial expansion, replacing
the Bloch plane wave state $\left|n{\bf k}\right\rangle $ with localized
site state $\left|{\bf r},\alpha\right\rangle $. The arithmetic and
geometric mean DOS for a disordered system are defined as $\rho_{\mathrm{a}}(\epsilon)=\left\langle \frac{1}{V}\sum_{i=1}^{V}\sum_{\alpha=1}^{4}\rho_{{\bf r}_{i},\alpha}(\epsilon)\right\rangle _{\mathrm{imp}}$
and $\rho_{\mathrm{t}}(\epsilon)=\exp\left[\frac{1}{N_{s}}\sum_{i=1}^{N_{s}}\left\langle \ln\sum_{\alpha=1}^{4}\rho_{{\bf r}_{i},\alpha}(\epsilon)\right\rangle _{\mathrm{imp}}\right]$,
respectively \cite{ZhangYY2012:PRB,ZhangYY2013:PRB}. Here we randomly
choose a finite number of lattice sites $N_{s}\ll V=L_{x}L_{y}L_{z}$
to improve the statistics of $\rho_{\mathrm{t}}$ \cite{Pixley2015:PRL}.
Other parameters are the same as the calculation of spectral functions.

\paragraph*{Normalized Localization Length}\label{subsec:NLL}

The normalized localization length serves as a fundamental metric
that underlies research on the physics of Anderson localization \cite{MacKinnon1983:ZPB,Ohtsuki2018:JPSP,Yamakage2013:PRB}.
In this approach, the disordered sample is fabricated into a lengthy
quasi-1D geometry along $x$ direction with finite cross-section $L_{y}\times L_{z}$,
and the Schr{\"o}dinger equation is rearranged into the following
form:
\begin{equation}\begin{aligned}
\left[\begin{array}{c}
\psi_{n+1}\\
\psi_{n}
\end{array}\right]=T_{n}\left[\begin{array}{c}
\psi_{n}\\
\psi_{n-1}
\end{array}\right] ,
\end{aligned}\end{equation}
where $\psi_{n}$ is the wavefunction of the $n$-th slice along the
longitudinal direction, and $T_{n}$ is the transfer matrix defined
as:
\begin{equation}\begin{aligned}
T_{n}=\left[\begin{array}{cc}
H_{n,n+1}^{-1}\left(E_{\rm F}-H_{n,n}\right) & -H_{n,n+1}^{-1}H_{n-1,n}^{\dagger}\\
I_{n} & 0
\end{array}\right] .
\end{aligned}\end{equation}
$H_{n,n}$ is the Hamiltonian of the $n$-th slice, and $H_{n,n+1}$
is the hopping matrix between the $n$-th and $\left(n+1\right)$-th
slice. $I_{n}$ is an identity matrix. Owing to the disorder, the
wave function decays exponentially as $\psi_{n,i}\sim\mathrm{e}^{\pm n/\lambda_{i}}$,
and the localization length $\lambda_{i}$ can be extracted as follows.
First, we define a consecutive product $O_{L_{x}}=\prod$$_{n=1}^{L_{x}}T_{n}$
that transforms the wave function of the first slice to the $L_{x}$-th
slice. According to the Oseledec's theorem, the limit
$O=\lim_{L_{x}\to+\infty}\left(O_{L_{x}}^{\dagger}O_{L_{x}}\right)^{\frac{1}{2L_{x}}}$
exists and its eigenvalues take on the form of $\left\{ \mathrm{e}^{\pm\gamma_{i}}\right\} $.
$\gamma_{i}$'s are called Lyapunov exponents. The smallest positive
Lyapunov exponent is found to be inversely related to the largest
localization length through the expression $\lambda_{x}=1/\gamma$.
Ultimately, the normalized localization length is defined as $\Lambda_{x}=\lambda_{x}/L_{y(z)}$,
depending on which quantity is changed. Furthermore, it is crucial
to introduce numerical stabilization after every few iterations of
transfer matrix multiplication, as accumulated round-off errors can
corrupt the calculation of Lyapunov exponents \cite{Yamakage2013:PRB}.

\paragraph*{Effective Medium Theory}\label{subsec:EMT}

To understand the origin and robustness of the HQHM in disorder,
we can employ the effective medium theory in conjunction with the
Kubo formula for electric conductivities \cite{Bastin1971:JPCS,ChenYu2018:PRB}.
The self-consistent Born approximation is a very useful tool to investigate
the physics in topological Anderson insulator \cite{Groth2009:PRL,Guo2010:PRL,Chen2015:PRL,Chen2017:PRB,Chen2018:PRB,Hua2019:PRB,Li2020:PRL,WangXR2020:PRR,Sarma2022:PRB,Chen2023:PRB}.
In the semi-magnetic structure with multiple layers, the electron
wave function has a layer degree of freedom, which lead to a matrix
structure of the Green's function and self-energy. We extend the self-consistent
Born approximation to the layered structure (see Supplementary Note
3 for details), within which the retarded self
energy is diagonal in the layer degree of freedom subspace and can
be expressed as: $\Sigma_{i_{z}i_{z}}^{R}(E_{\rm F})=\frac{W^{2}}{12S}\sum_{{\bf k}'}G_{i_{z}i_{z}}^{R}({\bf k}',E_{\rm F}),$
where $i_{z}$ is the layer number, $G^{R}(\mathbf{k}',E_{\rm F})=\left[E_{\rm F}-H_{\mathrm{0}}(\mathbf{k}')-\Sigma^{R}\right]^{-1}$
is the dressed Green's function. Despite the translational symmetry
breaking in the $z$ direction, our scheme is still applicable. The
self-consistent equations for self-energy can be solved numerically
\cite{Sarma2022:PRB}. With the help of the self-energy, the Hall
conductivity can be calculated by means of the Kubo-Bastin formula:
\begin{equation}\begin{aligned}
\sigma_{xy} & =\int_{-\infty}^{+\infty}\mathrm{d}\epsilon n_{\rm F}\int\frac{\mathrm{d}^{2}\mathbf{k}}{(2\pi)^{2}}
\mathrm{Tr}\left[\left(v_{x}\frac{\mathrm{d}G^{R}}{\mathrm{d}\epsilon}v_{y}-v_{y}\frac{\mathrm{d}G^{A}}{\mathrm{d}\epsilon}v_{x}\right)\left(G^{R}-G^{A}\right)\right] ,
\end{aligned}\end{equation}
where $G^{R}(\mathbf{k},\epsilon)=\left[\epsilon-H(\mathbf{k})-\Sigma^{R}(\mathbf{k},\epsilon)\right]^{-1}$.
Here, the self-energy $\Sigma^{R}(\mathbf{k},\epsilon)$ is approximated
as $\Sigma^{R}(\mathbf{k},E_{\rm F})$ that is numerically obtained in
the effective medium theory, and $v_{x/y}=\frac{\partial H}{\partial k_{x/y}}$
are the velocity operators. $n_{\rm F}=1/(\mathrm{e}^{(\epsilon-E_{\rm F})/T}+1)$
is the Fermi-Dirac distribution. In the effective medium theory, the
translational invariance in each layer is restored, allowing the momentum
$\mathbf{k}$ to be well-defined and conserved. $\mathrm{Tr}$ denotes
the trace, which is performed over the degrees of freedom associated
with layers, spin, and orbital.

\paragraph*{Quantum Interference Theory: Quantum Corrections to Conductivity}\label{subsec:QIT}

The quantum interference correction to conductivity arises from the
coherent backscattering of electrons due to impurity scattering, leading
to either WL or WAL, depending on the system's symmetries and the
nature of the scattering processes. To calculate the correction, the
diagrammatic approach has been developed to incorporate disorder-averaged
Green's functions, vertex corrections, and Cooperon modes. Consider
the low-energy effective Hamiltonian $H=H_{t}\oplus H_{b}$ which
describes the top ($H_{t}$) and bottom ($H_{b}$) surface states
of the system. A disorder potential including both intra-surface scattering
$U_{0}$ and inter-surface scattering $U_{1}$ is taken into account.
The disorder is modeled as short-range and characterized by the correlation
functions: $\langle U_{i}(\mathbf{r})U_{j}(\mathbf{r}^{\prime})\rangle_{dis}=U_{i}^{2}\delta_{ij}\delta(\mathbf{r}-\mathbf{r}^{\prime})$
where $i,j=0,1$ represent intra-surface and inter-surface scattering,
respectively, and $\langle...\rangle_{dis}$ denotes the disorder
averaging. The total quantum correction to conductivity, $\sigma_{qi}$,
is contributed by both intra-surface and inter-surface Cooperon channels.

\begin{figure}
\centering
\includegraphics[width=12cm]{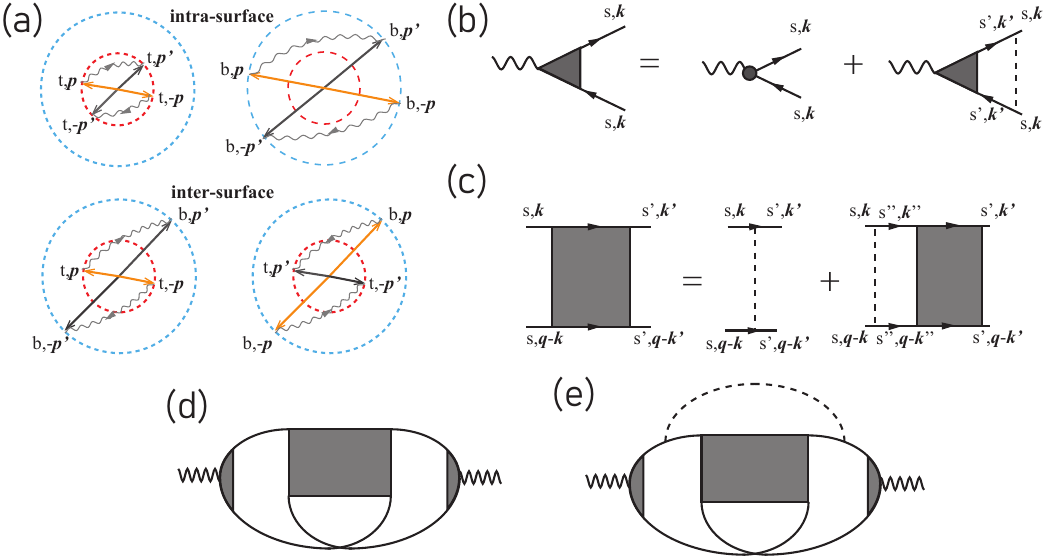}
\caption{\textbf{Feynman diagrams for evaluating quantum corrections to the longitudinal conductivity}.
(a) Diagrammatic representation of the bare scattering vertex for
intra- and inter-surface Cooperon channels. The blue and red circles
represent the Fermi surfaces for the bottom and top surface states,
respectively. Yellow and black arrows denote the incoming and outgoing
momenta. Diagrammatic representation of the self-consistent equation
for (b) the dressed vertex and (c) for the Cooperon. Leading-order
quantum correction to conductivity from (d) the bare Hikami box and
(e) the dressed Hikami box.}
\label{fig:Feynman_diagram}
\end{figure}

The intra-surface contribution is evaluated by means of the dressed
Hikami box formalism, which accounts for both bare and dressed vertex
corrections. The bare Hikami box represents the leading-order quantum
correction (Fig. \ref{fig:Feynman_diagram}(d)), while the dressed
ones (Fig. \ref{fig:Feynman_diagram}(e)) include higher-order corrections
arising from multiple scattering events. In systems with strong spin-orbit
coupling, such as topological insulators, the dressed Hikami box contribution
is significant and often carries an opposite sign relative to the
bare contribution, leading to a suppression or enhancement of the
quantum interference effect. The inter-surface contribution, on the
other hand, is influenced by the inter-surface scattering potential
$U_{1}$, which couples the top and bottom surface states. The contribution
is particularly important in thin films where the two surfaces are
in close proximity, allowing for significant inter-surface scattering.

The Cooperon functions are solved self-consistently via the Bethe-Salpeter
equation as illustrated in Fig. \ref{fig:Feynman_diagram}(c):
\begin{equation}\begin{aligned}
\Gamma_{ss^{\prime}}^{ss^{\prime}}(\theta_{\mathbf{k}},\theta_{\mathbf{k}^{\prime}},\mathbf{q})=\gamma_{ss^{\prime}}^{ss^{\prime}}(\theta_{\mathbf{k}},\theta_{\mathbf{k}^{\prime}})+&\int\frac{d^{2}\mathbf{k}^{\prime\prime}}{(2\pi)^{2}}\sum_{s^{\prime\prime}=t,b}\gamma_{ss^{\prime\prime}}^{ss^{\prime\prime}}(\theta_{\mathbf{k}},\theta_{\mathbf{k}^{\prime\prime}}) \times \\ 
& G_{s^{\prime\prime}}^{R}(\mathbf{k}^{\prime\prime})G_{s^{\prime\prime}}^{A}(\mathbf{q}-\mathbf{k}^{\prime\prime}) 
\Gamma_{s^{\prime\prime}s^{\prime}}^{s^{\prime\prime}s^{\prime}}(\theta_{\mathbf{k}^{\prime\prime}},\theta_{\mathbf{k}^{\prime}},\mathbf{q}) 
\label{eq:BS_equation}
\end{aligned},\end{equation}
where $s,s^{\prime}=t,b$ represent the top and bottom surfaces, $\theta_{\mathbf{k}}$
and $\theta_{\mathbf{k}^{\prime}}$ label the incoming and outgoing
momenta on Fermi surfaces $s$ and $s^{\prime}$, respectively, and
$\gamma_{ss^{\prime}}^{ss^{\prime}}$ is the bare scattering vertex
as illustrated diagrammatically in Fig. \ref{fig:Feynman_diagram}(a).
The Cooperon functions $\Gamma_{ss^{\prime}}^{ss^{\prime}}$ are categorized
into intra-surface modes (e.g., $\Gamma_{tt}^{tt}$ and $\Gamma_{bb}^{bb}$)
and inter-surface modes (e.g., $\Gamma_{tb}^{tb}$ and $\Gamma_{bt}^{bt}$).
The intra-surface Cooperon modes dominate when the scattering is confined
to one surface, while the inter-surface modes become relevant when
the inter-surface scattering is significant. The functions $\Gamma_{tb}^{bt}$
and $\Gamma_{bt}^{tb}$ describe the processes where the two outgoing
momenta lie on distinct Fermi surfaces. As a result, their total momentum
summation cannot vanish, making them irrelevant for the low-energy
Cooperon modes. After performing tedious calculations, as outlined
in Supplementary Note 4, the total quantum correction
to conductivity is expressed as:
\begin{equation}\begin{aligned}
\sigma_{qi}=\frac{e^{2}}{2\pi h}\left(\alpha^{t}\ln\frac{(l_{e}^{t})^{-2}+(l^{t})^{-2}}{L^{-2}+(l^{t})^{-2}}+\alpha^{b}\ln\frac{(l_{e}^{b})^{-2}+(l^{b})^{-2}}{L^{-2}+(l^{b})^{-2}}\right) ,
\end{aligned}\end{equation}
where the coefficients $\alpha^{t}$ and $\alpha^{b}$ are defined as
\begin{equation}\begin{aligned}
\alpha^{t}= & [(w_{0}^{t}+w_{2}^{t})(-1+\varpi_{tra,0}^{t})+w_{1}^{t}(1-\varpi_{tra,1}^{t})]\eta_{t}^{2}\\
 & +[w_{1}^{t}\varpi_{ter,1}^{t}-(w_{2}^{t}\varpi_{ter,2}^{t}+w_{0}^{t}\varpi_{ter,0}^{t})]\eta_{t}\eta_{b},\\
\alpha^{b}= & w_{1}^{b}(1-\varpi_{tra}^{b})\eta_{b}^{2}+w_{1}^{b}\varpi_{ter}^{b}\eta_{t}\eta_{b}.
\end{aligned}\end{equation}
The total quantum interference correction can be organized into two
distinct contributions, corresponding to the top ($t$) and bottom
($b$) surface states. Each of these contributions includes both intra-surface
and inter-surface Cooperon channel effects. Here, $w_{i}^{s}$ are
the weight factor for each Cooperon functions with angular momentum
$i=0,1,2$ and the surface $s=t,b$. $\varpi_{tra,i}^{s}$ and $\varpi_{ter,i}^{s}$
represent renormalization factors from the dressed Hikami box corrections
for intra- and inter-surface Cooperon channels, respectively. The
factors $\eta_{s}$ account for the vertex corrections, and $l^{s}$
is the characteristic length associated with each contribution, quantifying
the suppression of quantum interference due to the dephasing or symmetry-breaking
mechanism. The lengths play a crucial role in determining the strength
of the quantum correction, in which larger values of $l^{s}$ indicate
less suppression of quantum interference effects.

\backmatter

\bmhead{Data availability}
The data that support the findings of this study are available from the
authors upon reasonable request.

\bmhead{Code availability}
All numerical codes in this paper are available upon reasonable request to
the authors.

\bmhead{Acknowledgments}
We acknowledge Dr. Huan-Wen Wang and Rui Chen for
helpful discussions. This work was supported by the Research Grants
Council, University Grants Committee, Hong Kong under Grant Nos. C7012-21G
and 17301823, Quantum Science Center of Guangdong-Hong Kong-Macao
Greater Bay Area GDZX2301005, and Guangdong Basic and Applied Basic
Research Foundation No. 2024A1515010430 and 2023A1515140008.

\bmhead{Author contributions}
S.Q.S. conceived the project. S.H.B. performed the numerical simulations 
using the real-space Kubo formula, calculated the density of states and 
normalized localization length, and carried out the corresponding error analysis 
and effective medium theory analysis. F.B. performed numerical calculations of 
the spectral functions and Feynman diagrammatic calculations. 
All authors contributed to writing the manuscript and discussing the results.

\bmhead{Competing interests}
All authors declare no competing interests.

\bibliography{ref}

\end{document}